\documentclass[OSA,JOSAB,twocolumn]{revtex4}
\usepackage{amsmath}
\usepackage{amssymb}
\usepackage{graphics}
\begin{document}
\draft
\title{The Effects of Geometry on the Hyperpolarizability} 
\author{Mark G. Kuzyk}
\address{Department of Physics and Astronomy, Washington State University, Pullman,
Washington  99164-2814 \\ email: kuz@wsu.edu}
\author{David S. Watkins}
\address{Department of Mathematics, Washington State University, Pullman,
Washington  99164-3113 \\ email: watkins@math.wsu.edu}
\newcommand{\bibs}{c:/PCTeXv4/TexBiB/final}
\date{\today}
\begin{abstract}
Extensive studies in the past have focused on precise calculations of the nonlinear-optical susceptibility of thousands of molecules.  In this work, we use the broader approach of considering how geometry and symmetry alone play a role.   We investigate the nonlinear optical response of potential energy functions that are given by a superposition of force centers (representing the nuclear charges) that lie in various planar geometrical arrangements.  We find that for certain specific geometries, such as an octupolar-like molecule with donors and acceptors of varying strengths at the branches, the hyperpolarizability is near the {\em fundamental limit}.  In these cases, the molecule is observed to be well approximated by a three-level model - consistent with the three-level ansatz previously used to calculate the {\em fundamental limits}.  However, when the hyperpolarizability is below the {\em apparent limit} (about a factor of thirty below the {\em fundamental limit}) the system is no longer representable by a three-level model; where both two-level and a many-state models are found to be appropriate, depending on the symmetry. 
\end{abstract}
\maketitle

\section{Introduction}

Sum rules, as applied to the sum-over-states perturbation expression of the molecular nonlinear-optical susceptibilities,\cite{orr71.01} were first used to calculate the {\em fundamental limits} of the off-resonant hyperpolarizability $\beta$,\cite{kuzyk00.01,kuzyk03.02} and the second hyperpolarizability, $\gamma$.\cite{kuzyk00.02,kuzyk03.01} The role of off-diagonal elements, such as measured with Hyper Rayleigh Scattering,\cite{olbre20.01} were also considered.\cite{kuzyk01.01}  Indeed, the calculated {\em fundamental limits} were shown to be useful in culling out bad measurements.\cite{clays01.01} These calculations were later applied to calculating the resonant two-photon absorption cross-section\cite{kuzyk03.03,perez05.01} as well as the maximum possible two-photon absorption cross-section when a molecule is doubly-resonant with the excitation source.\cite{kuzyk04.02}  The theory is supported by the fact that no molecule has ever been found to break the {\em fundamental limit},\cite{Kuzyk03.04} it has been used as a guide to develop better small molecules,\cite{May05.01} and it is used to understand scaling laws.\cite{slepk04.01}

While all molecules ever measured are found to fall below the limit, it was pointed out that there is a gap between the {\em fundamental limit} and the molecules with the largest measured nonlinear-optical susceptibility.\cite{kuzyk03.02,Kuzyk03.05,kuzyk03.01}  This {\em apparent limit}, a factor of $10^{3/2}$ below the {\em fundamental limit}, has two possible implications.  Either, the theory could be flawed so that it overestimates the fundamental limit by more than a factor of thirty; or there is another factor that acts to suppress the nonlinear-optical response.  To test the hypothesis that the theory has overestimated the nonlinear response, all one needs to do is show that a system exists whose hyperpolarizability exceeds the {\em apparent limit}.  The clipped harmonic oscillator, whose hyperpolarizability can be calculated analytically without approximation, has a hyperpolarizability that is an order of magnitude larger than the {\em apparent limit}, yet about a factor of two below the {\em fundamental limit}.\cite{Tripa04.01} So, calculations of the {\em fundamental limit} do not appear to be flawed and seem to give a {\em fundamental limit} that is within reason.  Given that very talented organic chemists have been working almost three decades to improve the hyperpolarizability of molecules through structure-property studies, and no molecule has ever been found to breach the {\em apparent limit}, it is probably prudent to conclude that the best organic molecules fall below the {\em apparent limit} because of some inherent properties of the types of organic molecules that have been synthesized.

While nanoengineering methods have been used to increase the second hyperpolarizability per molecule to within a factor of two of the {\em fundamental limit},\cite{wang04.01} a more careful analysis shows that the interactions between molecules makes them respond collectively.  As such, the collection of molecules is acting like a single supermolecule whose response is still well below the {\em apparent limit}.  So, the state of affairs with regards to molecular hyperpolarizabilities is that all molecules ever measured and all analytical calculations ever performed fall below the {\em fundamental limit}, though some of the calculations come close.  This set of evidence supports the theory of {\em fundamental limits} and the use of the three-level ansatz.\cite{kuzyk05.01,kuzyk05.02a}  So, while a three-level model implies truncation of the sum rules, which in the Sum Over States (SOS) expression for $\beta$ could lead to pathologies, the fact that the results of this process are consistent with observation suggests that they are correct, though these nuances deserve further study.

In the present work, we investigate how the geometry of a molecule affects its off-resonant hyperpolarizability and conditions that lead to the maximum response.  In particular, we consider how molecular properties - such as energy-level spacing and transition moments - are affected by symmetry and set conditions that lead to a maximized response that breaches the {\em apparent limit} and approaches the {\em fundamental limit}.  We find that at the {\em fundamental limit}, the system becomes a three-level system; therefore, the three-level ansatz is obeyed.

\section{Theory}

\subsection{Fundamental Limits}

The truncated sum rules have been shown to lead to an absolute limit that scales properly when compared with the set of {\em all measured values of $\beta$}.\cite{kuzyk00.01,kuzyk03.02}  In particular, the very best molecules fall on a curve that is a factor of $10^{-3/2}$ below the {\em fundamental limit}.  Even so, there have been questions raised as to the validity of this approach\cite{champ05.01} since the truncation process leads to pathologies in some of the resulting sum rule equations.  However, the pathological equations can be eliminated based on the physical principles that the resulting theory should be self-consistent, consistent with observations, and free from divergences.\cite{kuzyk05.01}

The three-level model is not meant to predict $\beta$ for the general molecule with many states, but to treat a molecule that is getting close to being in the near-ideal state of maximized $\beta$ by concentrating, in correct proportion, all of the oscillator strength into the two dominant excited states.  Then, by treating the system as a three-level model (under these circumstances, the truncated sum rules are an excellent approximation), we can algebraically maximize the mathematical expression to investigate what combinations of moments will bring us to the maximum.  We thus assume the proposed ansatz that any quantum system close to the {\em fundamental limit} is represented as a three-level model and that the truncated sum rules, reduced in number by the physical requirements listed above to the reduced set of sum rules, thus yield the correct limiting case.  The calculations we present here are accurate enough to test the validity of this approximation.

The process of applying the sum rules to the three-level model of the sum-over-states expression of the hyperpolarizability, $\beta$, is described in the literature.\cite{kuzyk00.01,kuzyk03.01,kuzyk00.02,kuzyk03.02}  For $\beta$, this yields,
\begin{equation}
\beta = 6 \sqrt{\frac {2} {3}} e^3 \frac {\left| x_{10}^{MAX} \right|^3} { \sqrt[4]{3} E_{10}^2} G(X) f(E) = \beta_0 G(X) f(E) ,\label{betafG}
\end{equation}
where
\begin{equation}
f(E) = (1-E)^{3/2} \left( E^2 + \frac {3} {2} E + 1 \right),\label{DEFf(E)}
\end{equation}
\begin{equation}
G(X) = \sqrt[4]{3} X \sqrt{\frac {3} {2} \left( 1 - X^4\right)},\label{defG(X)}
\end{equation}
\begin{equation}
X = \frac {x_{10}} {x_{10}^{MAX}} , \label{Xfrac}
\end{equation}
\begin{equation}
E = \frac {E_{10}} {E_{20}} ,\label{Efrac}
\end{equation}
and
\begin{equation}
x_{10}^{MAX} = \sqrt{ \frac {\hbar^2 N} {2mE_{10}} } .  \label {xsum}
\end{equation}

Assuming that $E$ and $X$ are independent,  $f(E)$ peaks at $E=0$ and $G(X)$ peaks at $X=\sqrt[-4]{3}$.  The maximum value of each function is unity, so from Equation \ref{betafG}, we get the {\em fundamental limit}
\begin{equation}
\beta_{MAX} = \beta_0 f(0) G(\sqrt[-4]{3}) =  \sqrt[4]{3} \left( \frac {e \hbar} {\sqrt{m}} \right)^3 \left[ \frac {N^{3/2}} {E_{10}^{7/2}} \right] . \label{betaMAX3L}
\end{equation}

It is important to point out that because of the sum rules, which can be written in general form as
\begin{equation}
\sum_{n=0}^{\infty} \left( E_n - \frac {1} {2} \left( E_l + E_p \right) \right) x_{ln} x_{np} = \frac {\hbar^2 N} {2m} \delta_{l,p},
\label{sumrule}
\end{equation}
the sum rule equation with $(m,p)=(0,0)$, truncated to three levels, yields,
\begin{equation}
\left| x_{10} \right|^2 + \left| x_{20} \right|^2 E^{-1}= \frac {\hbar^2} {2mE_{10} } N .
\label{groundsumrule(3)}
\end{equation}
As such, since $0 \leq X \leq 1$, if $X=1$, all of the oscillator strength is focused into the first excited state; but, when $X=0$, the second excited state gets all of the oscillator strength.  So, this expression is appropriate for the three-level ansatz, where both states will contribute.

\subsection{Numerical Techniques}

Our work focuses on understanding how the symmetries of one- and two-dimensional molecules affect the hyperpolarizability.  As such, we solve the two-dimensional Schr\"{o}dinger eigenvalue problem
\begin{equation}\label{eq:schroedinger}
-\frac{\hbar^{2}}{2m} \nabla^{2}\Psi + V\Psi = E \Psi
\end{equation}
for the lowest ten to 25 energy levels, depending on the degree of convergence of the resulting hyperpolarizability.  Since our problem is in 2D, we use a logarithmic potential.  For $k$ nuclei with charges $q_{1}e$, \ldots, $q_{k}e$ located at points $s^{(1)}$, \ldots, $s^{(k)}$, the potential is 
$$
V(s) = \frac{e^{2}}{L}\sum_{j=1}^{k}q_{j}\log \|s - s^{(j)} \|,
$$
where $L$ is a characteristic length.  We assume $L = 2 \AA$, which has
the effect that the force due to a charge at distance $2 \AA$ is the
same as it would be for a Coulomb potential.

We discretize the eigenvalue problem given by Equation \ref{eq:schroedinger} using a
quadratic finite element method \cite{zienk05.01,atkin01.01} and solve the
resulting matrix eigenvalue problem for the ten to 25 smallest eigenvalues
and corresponding eigenvectors by the implicitly-restarted Arnoldi
method \cite{soren92.01} as implemented in ARPACK \cite{lehou98.01}.  
Each eigenvector yields a wave function $\Psi_{n}$ corresponding to
energy level $E_{n}$.  The moments 
$$x_{mn} = \int_{-\infty}^{\infty}\!\!\int_{-\infty}^{\infty} s_{1}
\Psi_{m}(s_{1},s_{2}) \Psi_{n}(s_{1},s_{2})\, ds_{1}ds_{2}$$
are computed, and these and the energy levels $E_{n}$ are used to
compute $\beta$ and all other quantities by the formulas in the literature.\cite{Tripa04.01}

\section{Results and Discussion}

In this section, we show how the hyperpolarizability depends on symmetry by considering various arrangements of force centers in two-dimensions.  In addition, these studies can be used to determine which geometries lead to enhanced hyperpolarizability.  We will show how this process sheds light on why there is a large gap between the best molecules and how one can approach the {\em fundamental limit}.  Geometries that we will consider include:

\begin{itemize}  
\item Two force centers of varying charge asymmetry and separation, which tests how $\beta$ depends on the dipole moment.
\item Three force centers on a circle of varying charge asymmetry and angular separation, which tests how $\beta$ varies as the systems makes a transition from a monopole, to an octupolar arrangement to a dipole, including the continuum between these limits.
\item Three force centers on a circle with a fourth charge at the center, where the middle force center's charge is varied.
\end{itemize}

Since the purpose of these calculations is to study only the effects of symmetry, all other details of the system must be suppressed.  (This is in contrast to most real molecular structures, in which it is difficult to separate symmetry from other effects.)  To this end, we assume that the net nuclear charge is $+e$ and that a single electron of charge $-e$ is placed into this potential.  As such, we can imagine the nuclear skeleton as a collection of force centers that are surrounded by tightly-bound electrons, which screen the nuclei leaving a net charge of $+e$.  For simplicity, we assume these nuclear centers and screening electrons are point charges, and we calculate the wavefunction of the remaining electron with no electron correlations, from which we get the hyperpolarizability.

\subsection{Dipolar Symmetry}

\begin{figure}
\centering
\includegraphics{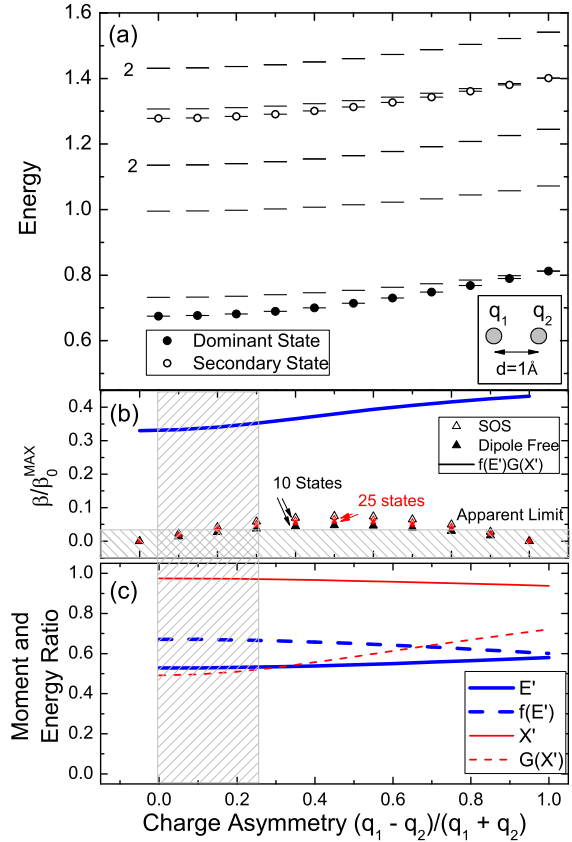}
\caption{(a) The calculated excited-state energy levels relative to the ground state (with degenerate states labelled by their degeneracy), where the closed circles label the states with the largest transition moment to the ground state and the open circles represent the states with the second largest transition moment; (b) hyperpolarizabilities, where the vertical shaded box represents the charge asymmetry range of typical organic molecules and the horizonal shaded region represents the area within the {\em apparent limit horizon}; and (c) relevant normalized parameters as a function of charge asymmetry between two charges separated by a distance of 1$\AA$.\label{fig:2charge_asymmetry1}}
\end{figure}

Figure \ref{fig:2charge_asymmetry1} shows the calculated energy levels, hyperpolarizability, and the quantities defined in Equations \ref{betafG}, \ref{DEFf(E)}, \ref{defG(X)}, \ref{Xfrac}, and \ref{Efrac} for a diatomic molecule as a function of charge asymmetry.

There are two equivalent sum-over-states (SOS) expressions that can be used to calculate $\beta$.  The standard expression is overspecified since it is possible to pick transition moments and energies that are inconsistent with the sum rules, therefore violating the Schr\"{o}dinger Equation.  Using the sum rules, the dipole terms in the SOS expression can be removed, yielding a more simple equation.\cite{kuzyk05.01a,kuzyk05.02}  The SOS results are shown with open triangles and the dipole-free expression is given by the solid triangles.  The larger triangles represent calculations using 10 states (the energy levels of the excited states are shown in the upper part of Figure \ref{fig:2charge_asymmetry1}) while the smaller squares show calculations using 25 states.  The difference between the two can be used as a test of convergence of the SOS expression.\cite{kuzyk05.02}  Note that all values are normalized to the {\em fundamental limit} and all excited-state energies are relative to the ground state energy.

There are several conclusions that we can draw from the calculated values of $\beta$.  First, the largest hyperpolarizability, for a charge asymmetry of $(q_1 - q_2)/(q_1 + q_2) = 0.55$ is still more than an order of magnitude smaller than the {\em fundamental limit}.  Also, even at 25 states, the calculation has still not fully converged.  (Both the SOS and dipole-free expression should yield the same value at convergence.)  Secondly, we note that the separation between sites used in this calculation is comparable to the separation between atoms in an organic molecule; and, the shaded region with a charge asymmetry of $0.0 \leq \epsilon \leq 0.25$ is the typical range for organic molecules.  In the shaded region, the largest value of $\beta$ is less than about $0.04$, consistent with the largest measured values of $\beta$ and the {\em apparent limit}(horizontal shaded region).  As such, we find that for parameters typical of an organic molecule, we get a small value of $\beta$; and, we find that even for 25 states, the calculation has not yet converged, implying that many excited states contribute to $\beta$.  So, our 2-D calculations are at least approximately consistent with the behavior of real molecules.

When analyzing the importance of the excited state energies and transition moments, it is more useful to reexpress Equations \ref{betafG}, \ref{DEFf(E)}, \ref{defG(X)}, \ref{Xfrac}, and \ref{Efrac} so that $E_{10}$ is the excitation energy to the state with the largest transition moment and $E_{20}$ is the excitation energy of the state with the second largest transition moment.  As such, when the molecule approaches the {\em fundamental limit}, these are the two states that share the total oscillator strength of the molecule.  Thus, we define $E'$ and $X'$ to be the parameters given by Equations \ref{Xfrac} and \ref{Efrac} when the two states of interest are the ones with the largest oscillator strengths while the unprimed parameters refer to the two states with lowest excitation energies.  The solid circles in Figure \ref{fig:2charge_asymmetry1}a label the states with the largest transition moment to the ground state while the open circles label the second-most important state.  In Figure \ref{fig:2charge_asymmetry1}b, the solid line represents the function $f(E')G(X')$.  The data points representing the calculated values of $\beta$ of the model molecule, however, are normalized to the {\em fundamental limit}, which uses Equation \ref{betaMAX3L} with $E_{10}$ as the state with lowest excitation energy.

According to the three-level ansatz, near the {\em fundamental limit}, the quantum system should be approximated by a three-level model.  Since many states are required to calculate $\beta$, clearly, the diatomic molecule with a $1 \, \AA$ bond length is far from the {\em fundamental limit}.  This is consistent with the fact that the function $f(E')G(X')$ (solid line in Figure \ref{fig:2charge_asymmetry1}b) is well above the data (triangles, in Figure \ref{fig:2charge_asymmetry1}).  Both $f(E')>0.5$ and $G(X')>0.5$, so $X'$ and $E'$ of the two dominant excited sates are close to optimal.  Since $\beta$ is far from optimized, this would suggest that many excited states contribute to $\beta$, thus diluting the response.

\begin{figure}
\centering
\includegraphics{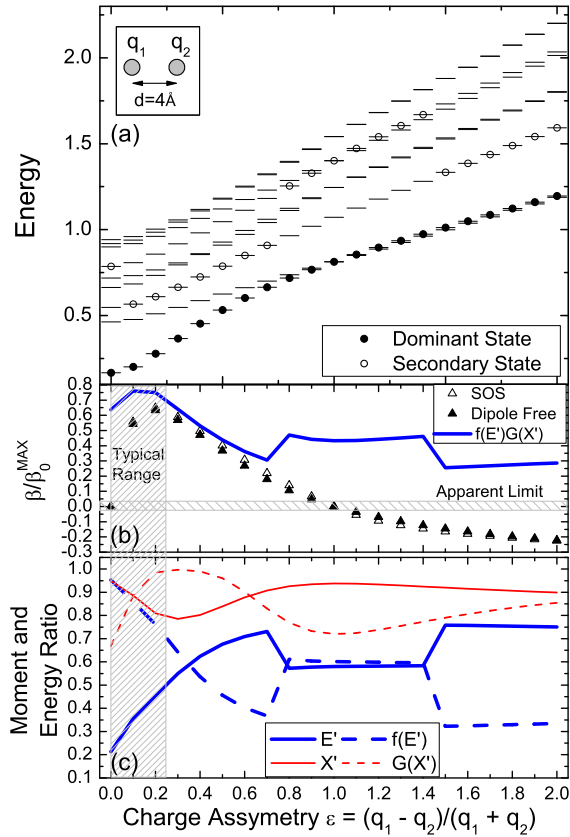}
\caption{(a) The calculated energy levels, where the closed circles label the states with the largest transition moment to the ground state and the open circles represent the states with the second largest transition moment; (b) hyperpolarizabilities, where the vertical shaded box represents the charge asymmetry range of typical organic molecules and the horizonal shaded region represents the {\em apparent limit}; and (c) relevant normalized parameters as a function of charge asymmetry between two charges separated by a distance of 4$\AA$.  Note that $q_1 + q_2 = 1e$ for all cases.\label{fig:2charge_asymmetry4}}
\end{figure}
Figure \ref{fig:2charge_asymmetry4} shows the same calculation with two force centers separated by $4 \AA$ for 10 states.  Note that even for this small number of states, the calculation has almost converged.  Also, $\beta$ is well above the {\em apparent limit} and is approaching the {\em fundamental limit}.  As such, while the system is not exactly a 3-level model, far fewer states are required than for the case shown in Figure \ref{fig:2charge_asymmetry1}.  Indeed, at the peak, the two calculations of $\beta$ yield the smallest fractional deviation, showing the highest degree of convergence.  This is consistent with the ansatz that as fewer excited states contribute, the system more closely approaches the ideal for maximizing $\beta$.  Furthermore, the function $G(X')f(E')$ is a better approximation to the calculated values of $\beta$ at the maximum, and, the energy spacing, quantified by the function $f(E')$ and the oscillator strength, quantified by $G(X')$ are both nearly maximized.  Indeed, the plot of the energy levels clearly shows how the first two excited states are well separated when $\beta$ approaches the limit.  Note that while the two dominant states have the largest separation for zero charge asymmetry, $\beta = 0$ in the centrosymmetric limit, as required by symmetry arguments for an even-order susceptibility.

Since the calculated value of $\beta$ varies slowly beyond a charge asymmetry of $2.0$, our plot does not extend into the slowly-varying regime.  However, it is important to note that at a charge asymmetry of about $5.3$, the negative value of normalized $\beta$ peaks at about $-0.305$ and approaches zero as the charge asymmetry is further increased.

Our numerical calculations are in accord with the assumptions used in calculating the {\em fundamental limits} and with the ansatz that $\beta$ is maximum when the system collapses into a three-level model.  Furthermore, $\beta$ vanishes in the centrosymmetric limit, as expected for an even-order susceptibility, and when $\epsilon = 1$, when one of the charges vanishes, again leaving a centrosymmetric system.  In the highly polar limit, $\beta$ vanishes, which is consistent with both a vast body of literature and the sum rules, which show that an arbitrarily large dipole moment difference necessarily is accompanied by a small transition moment to that state.\cite{kuzyk05.02a}  Most importantly, $\beta$ for a physically reasonable amount of charge asymmetry peaks near the {\em fundamental limit}.  However, the nuclear separation of $4 \AA$ is much larger than for typical organic molecules, in which the separation between carbon atoms is between about $1 \AA$ and $1.5 \AA$.  This may be one factor that contributes to the thirty-fold gap between the {\em fundamental limit} and the best 1-dimensional organic molecules.

\subsection{Octupolar Symmetry}

\begin{figure}
\centering
\includegraphics{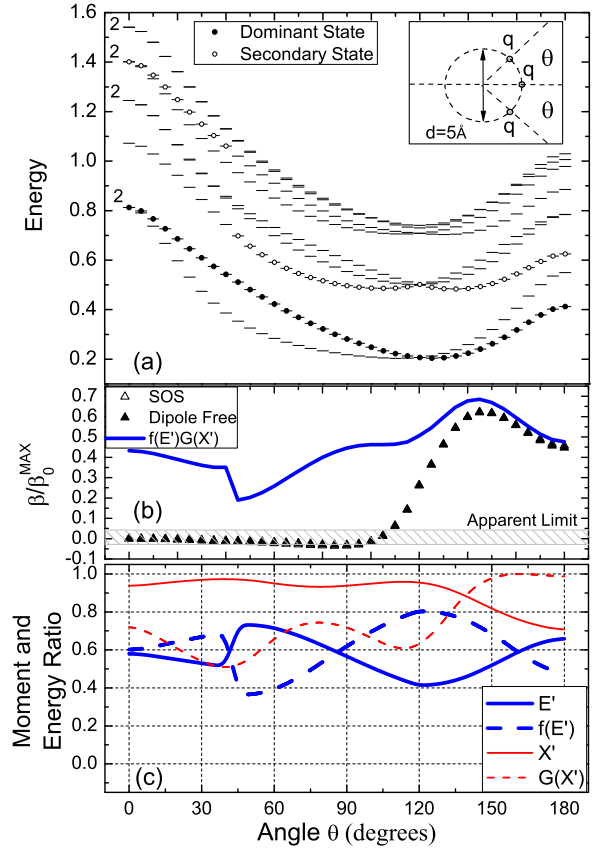}
\caption{(a) The calculated energy levels (with degenerate states labelled by their degeneracy), where the closed circles label the states with the largest transition moment to the ground state and the open circles represent the states with the second largest transition moment; (b) hyperpolarizabilities, where the horizonal shaded region represents the {\em apparent limit}; and (c) relevant normalized parameters as a function of angle between two of the charges in a three equal-charge system where each charge is q=1/3.\label{fig:3charge_theta}}
\end{figure}
Figure \ref{fig:3charge_theta} shows the diagonal component of the hyperpolarizability along the horizontal axis of a molecule of three equal charges (each of magnitude e/3) confined to a circle of diameter $5\AA$, as shown in the inset of Figure \ref{fig:3charge_theta}a.  One charge is fixed and the other two charges are symmetrically moved around the circle.  When $\theta = 0$, the charge distribution forms a monopole while for $\theta = 120^o$, the system is a pure octupole (i.e. when monopole and dipole moments vanish).  When $\theta = 180^o$ the molecule has strong dipole character.  As such, the system goes from monopole $\rightarrow$ octupole $\rightarrow$ dipole as $\theta$ varies continuously from $0^o$ to $180^o$.

The hyperpolarizability depends on angle as expected.  For example, several cases of energy-level crossing and degeneracies are observed at $\theta=120^o$, as expected for states near the ground state for a system with three-fold symmetry.  Also, while only 10 states are used in the calculation, the SOS expression and dipole-free expression - within the resolution of the size of the triangles, both yield the same hyperpolarizability; an indication that the calculation has converged.  This is consistent with the fact that the oscillator strength is shared among a small number of states so that the largest observed value of $\beta$ at around $145^o$ is near the {\em fundamental limit}.

A molecule has a nonzero even-order nonlinear-optical response only if it is non-centrosymmetric.  For a second-order process, only polar or octupolar symmetries will yield a non-vanishing hyperpolarizability.\cite{Joffr92.01} An interesting question is which symmetry yields the largest $\beta$.  For the purely octupolar molecule, when $\theta = 120^o$, the hyperpolarizability is well above the {\em apparent limit} with $\beta = 0.3 \beta_0^{MAX}$.  However, the hyperpolarizability peaks at $\theta \approx 145^o$ with $\beta \approx 0.6 \beta_0^{MAX}$.  So, we conclude that the best molecular symmetry is a mixture of octupolar and dipolar character.  In the monopole regime, $\beta$ is negative and below the {\em apparent limit}.  It becomes large and positive only as the system approaches the octupolar limit.

At $120^o$, when the molecule is purely octupolar, the function $f(E')$ is near a maximum because the two dominant states are the furthest apart in energy.  It is interesting to note that the first two excited states become degenerate, but only one of those degenerate states dominates in its oscillator strength.  In fact, at that point, the energy of the dominant state crosses over from being the second excited state for $\theta<120^0$ to being the lowest energy first excited state for $\theta>120^0$.  Interestingly, $\beta$ is by far the largest - and well above the {\em apparent limit}, when the dominant state is of lowest energy.  Most importantly, near the peak of maximum calculated $\beta$, the value of $f(E')G(X')$ and $\beta$ are nearly equal, which again shows that the three-level ansatz holds for molecules whose susceptibilities are near the {\em fundamental limit}.

Finally, it is interesting to note that the the function $G(X')$ is near unity for all angles larger than about $150^o$.  In this regime, the decrease in $\beta$ can be attributed to $f(E')$ since the two dominant state energies are getting closer together.  At $\theta = 180^o$, the three-charge system reduces to the diatomic molecule with a charge asymmetry of 1/3.  Comparing these results with Figure \ref{fig:2charge_asymmetry4}, where $G(X')$ is near unity between a charge asymmetry of 0.2 and 0.5, the value of $\beta$ in the 2-charge system is also clearly dominated by $f(E')$.  A comparison of these two cases also shows that for a diatomic system, a charge separation of $4 \AA$ is near optimum, giving a larger value of $\beta$ than for a $1 \AA$ or $5 \AA$ separation.

\begin{figure}
\centering
\includegraphics{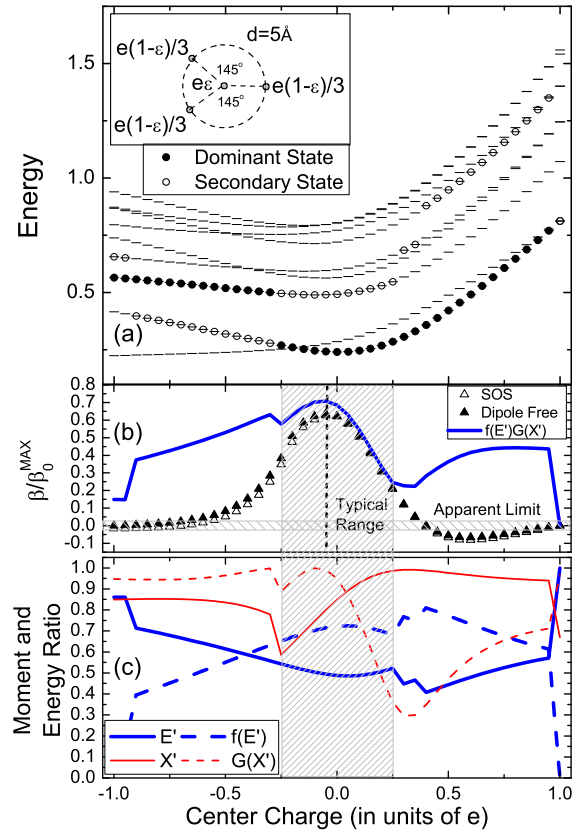}
\caption{(a) The calculated energy levels, where the closed circles label the states with the largest transition moment to the ground state and the open circles represent the states with the second largest transition moments; (b) hyperpolarizabilities, where the horizonal shaded region represents the {\em apparent limit} (the dashed vertical line shows the peak of $\beta$); and (c) relevant normalized parameters as a function of the amount of charge added to the center of the three-charge system fixed at $145^o$.  The three charges on the circle of radius $5 \AA$ are the same and the total charge of the 4-charge system is $+e$.\label{fig:vary_center_charge_angle145}}
\end{figure}

The charge arrangement in Figure \ref{fig:3charge_theta} is unphysical when compared with typical octupolar molecules, which are usually conjugated from the center outwards along the three prongs of the molecule.  To make the system more realistic, we have fixed the angle between the charges at $145^o$, which yields the largest Figure $\beta$ but have added a center charge to represent conjugation through the center, as shown in the inset of Figure \ref{fig:vary_center_charge_angle145}.  The hyperpolarizability along the horizontal axis of such a molecule of diameter $5\AA$ is calculated and plotted in Figure \ref{fig:vary_center_charge_angle145}b as a function of the center charge under the constraint that the total charge is fixed at $+e$.  When the center charge is $-0.05e$, normalized $\beta$ is maximized.  This suggests that a small amount of negative charge in the center of the molecule acts to push the electrons to the outer parts of the molecule, which appears to be advantageous for maximizing $\beta$.

At a center charge of $+1e$, all of the charges on the circle vanish, making the molecule centrosymmetric resulting in a vanishing value of $\beta$.  Similarly, when the center charge is about $-e$, $\beta$ also becomes small.  Note that for a center charge of $-e/4$, the second excited state becomes dominant and the first excited state becomes of secondary importance.  This inversion is only observed in this one geometry.  However, the calculated value of $\beta$ continues to behave smoothly.

\begin{figure}
\centering
\includegraphics{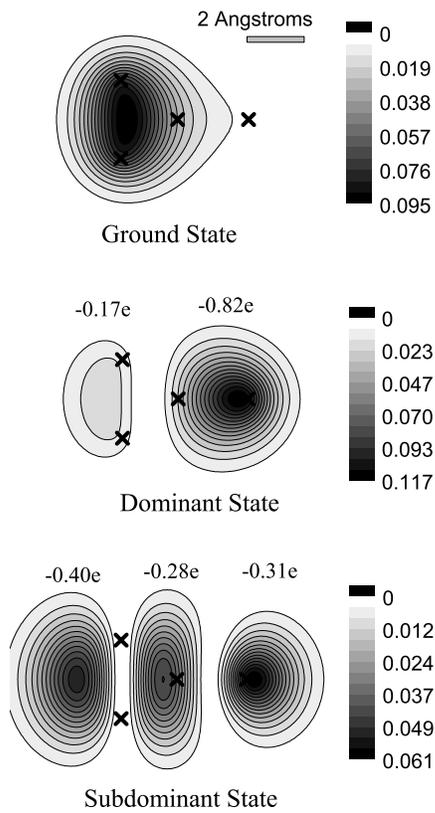}
\caption{Contour plots of the electron density for the ground state and the two dominant excited states for the molecule shown in the inset of Figure \ref{fig:vary_center_charge_angle145} with a center charge of $-0.05 \, e$.  The values displayed above the plots represent the total charge in that part of the wavefunction.  The ``$\times$" symbols represent the positions of the nuclei.\label{fig:WaveFunctionCountourAll}}
\end{figure}

Figure \ref{fig:WaveFunctionCountourAll} shows a plot of the electron density in the ground and two dominant excited states for the molecule shown in the inset of Figure \ref{fig:vary_center_charge_angle145} with a center charge of $-0.05 \, e$.  The electron in the ground state is mostly shared by the two charges on the left.  The transition from the ground state to the dominant state shows strong charge transfer yielding a dominant excited state with $82 \%$ of the charge centered on the righthand charge.  This is a strong dipolar transition with a large oscillator strength.  The higher-energy sub-dominant state, on the other hand, shows almost an equal distribution of charge across the whole molecule.  Our calculations thus suggest that the ideal molecule may be an octupolar one with slightly broken symmetry that results in strong but asymmetric charge localization. 

It is useful to consider in some detail the properties that lead to such a large response since this paints a portrait of the ideal molecule with a $\beta$ value that approaches the {\em fundamental limit}.  This hybrid molecule with optimum geometrical and dynamical parameters that lead to a maximum value of $\beta$, then, has two equal donor groups on one side of the molecule and an acceptor group on the third prong lies along the symmetry axis between the two donors.  The force center at the geometric center of the molecule is slightly negatively charged.  Due to the asymmetry of this geometry along the excitation axis, all of the lower-energy states are non-degenerate and two well-separated excited states dominate the response.  Furthermore, these two excited states share oscillator strength in just the right delicate proportion as confirmed by the fact that $G(X) \approx 1$.  Interestingly, the electron density is well localized on the two donor groups in the ground state and centered on the acceptor in the dominant excited state.  On the other hand, the subdominant state shows full charge delocalization.  There is no reason why such a molecule could not be synthesized.

\begin{figure}
\centering
\includegraphics{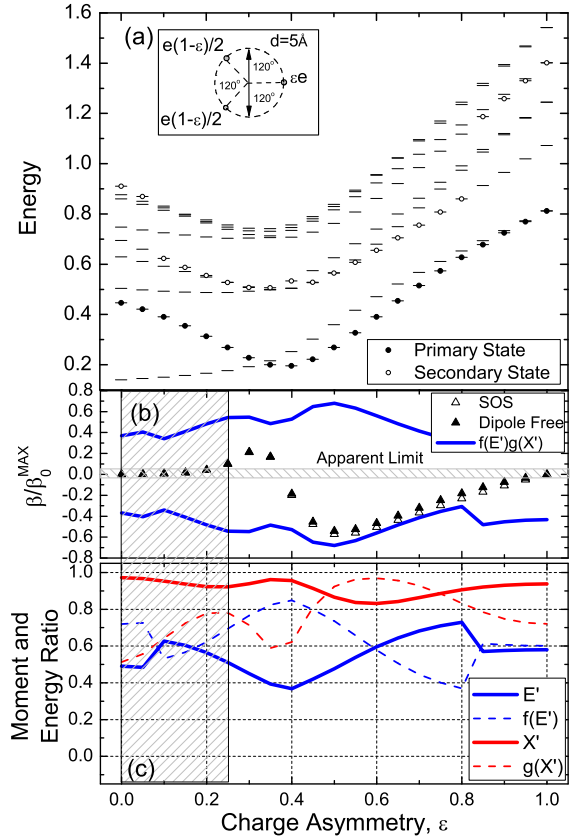}
\caption{(a) The calculated energy levels, where the closed circles label the state with the largest transition moment to the ground state and the open circles represent the states with the second largest transition moments; (b) hyperpolarizabilities, where the horizonal shaded region represents the {\em apparent limit}; and (c) relevant normalized parameters as a function of charge asymmetry between the charge on the left and the two charges on the right where the three charges are equally spaced at $120^o$.\label{fig:3charge_assymetry5-120-prime}}
\end{figure}

From the perspective of designing and synthesizing a molecule, it is useful to have other means of controlling the symmetry in addition to varying the angle of the bonds.  Another option is to fix the positions of the three force centers, and allow the magnitude of the charges to vary.  Since it is possible to make planar octupolar molecules with various donor and acceptor groups on their three ends, we calculate the case of three evenly-spaced charges on a circle of diameter $5 \, \AA$.  Two of the force centers have equal charges (each of magnitude $e(1-\epsilon)/2$), with the $\beta$ component measured along the symmetry axes running through the unique charge (of magnitude $\epsilon e$) and bisecting the line between the other two charges, as shown in the inset of Figure \ref{fig:3charge_assymetry5-120-prime}.  As the parameter $\epsilon$ is varied from $0$ through $1$, the net core electron charge varies from being evenly split between the left and righthand side of the molecule to all being on the left.  When $\epsilon = 1/3$, the core electrons are evenly shared by the three nuclei leading to a nuclear charge distribution that forms a pure octupole.

For the pure octupole, when all three charges are equal ($\epsilon = 1/3$), $\beta$ is a local maximum; and, its value is consistent with Figure \ref{fig:3charge_theta} when $\epsilon=0$.  Up to the octupolar limit, the second excited state is the dominant one.  For larger charge asymmetry beyond the observed crossover point between the two lowest-energy states, where the sign of $\beta$ changes, the first excited state dominates and $\beta$ peaks at a negative value when $\epsilon = 1/2$.  At the peak, then, there is an equal amount of charge on the left and right sides of the molecule.  So, by controlling the relative strengths of the charges between the two sides of the molecule to balance the charge, the optimum is reached.

Once again, near the peak value of $\beta$ - which is near the {\em fundamental limit}, the measured values of $\beta$ are near $f(E')G(X')$ and the second most dominant state is the third excited state.  $\beta$ is near $f(E')G(X')$ until a charge asymmetry of about $\epsilon = 0.8$, at which point the sixth excited state becomes the second most dominant state.  Above $\epsilon = 0.8$, $\beta$ is much smaller than $f(E')G(X')$.  So, the simplest paradigm for reaching the {\em fundamental limit} may be to synthesize molecules with three-fold symmetry with donors and acceptors of unequal strength.

Recall that when the {\em fundamental limit} of a nonlinear-optical susceptibility is calculated, the three-level ansatz is used; that is, at the {\em fundamental limit} the system must be identically a three-level system.  While the present calculations show some large values of $\beta$ that approach the {\em fundamental limit},  the best molecules are still more than a factor of 1/3 from the limit.  So, it is interesting to check if a three-level system appears to be a good approximation when its hyperpolarizability is close to the {\em fundamental limit}.
\begin{figure}
\centering
\includegraphics{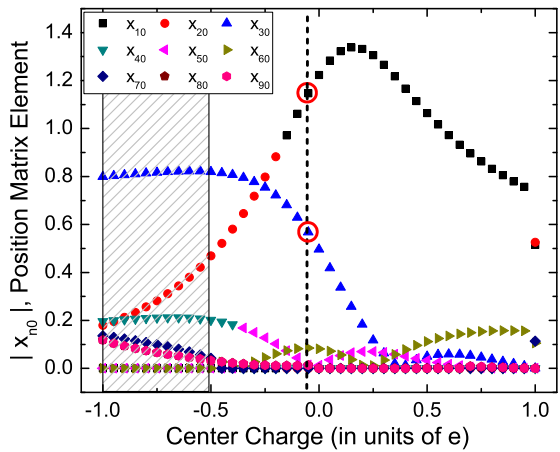}
\caption{Transition moments from the ground state as a function of the center charge for the molecule described in Figure \ref{fig:vary_center_charge_angle145}.  The shaded region corresponds to the range of center charge that yields a value of $\beta$ that is within the {\em apparent limit}.\label{fig:ansatz}}
\end{figure}

Figure \ref{fig:ansatz} shows the magnitude of the transition moments from the ground state as a function of center charge for the molecule described in Figure \ref{fig:vary_center_charge_angle145} for the lowest nine energy levels.  The dashed vertical line shows the value of charge asymmetry at which the hyperpolarizability peaks.  The circled points correspond to the two largest transition moments.  Note that Figure \ref{fig:vary_center_charge_angle145} shows that nine states are sufficient based on the convergence test of $\beta$.\cite{kuzyk05.02}  As such, we can conclude that only two excited states dominate the hyperpolarizability, consistent with the ansatz that when a quantum system has a hyperpolarizability at the {\em fundamental limit}, the system is identically a three-level system.  In our calculation, however, it must be pointed out that $\beta$ is not at the {\em fundamental limit} and while two states dominate, there are still small contributions from a third excited state, which is probably the cause of $\beta$ falling shy of $f(E')G(X')$.  Also, we should point out that for a central charge of about 0.4, only one excited state dominates so that the system is represented by a two-level model.  However, $\beta$ at that point is small, and is near the {\em apparent limit}.  This illustrates how a two-level model (which is commonly used for analysis by experimentalists) is not useful in describing systems near the {\em fundamental limit}, and is also consistent with the three-level ansatz.

The shaded region in Figure \ref{fig:ansatz} corresponds to a range in the center charge's magnitude that places the hyperpolarizability below the {\em apparent limit}.  In this region, the system is clearly no longer described by a three-level model.   This is consistent with the fact that all molecules ever measured fall below the {\em apparent limit} and that most molecules are not well described by a three-level model.  Indeed, the cause of the gap between the {\em apparent limit} and the {\em fundamental limit} is most likely due to the fact that the oscillator strength in most real molecules is spread over many excited states, diluting the nonlinear-optical response.  The trick, then, is finding molecules that have only two dominant excited states.

\section{Conclusion}

We have calculated effects of symmetry and geometry on the off-resonant hyperpolarizability of two-dimensional systems and find that the results are near the {\em fundamental limit} under certain conditions.  For example, the hyperpolarizability peaks at a negative value in an octupolar molecule when the three force centers have charges $e/4$, $e/4$, $e/2$, corresponding to a molecule with two donors on one side of the molecule each of half the strength as the one acceptor on the other side.  On the other hand, the hyperpolarizability peaks at a positive value when two of the three prongs are separated by $145^o$; and, a fourth charge, when placed at the center that acts as a donor increases $\beta$.  In this case, the ground state and dominant exited state are strongly localized and are connected by a large-oscillator-strength dipolar transition.  This may perhaps be the ideal moelcule.

In those cases where the hyperpolarizability peaks near the {\em fundamental limit}, the system is found to be well approximated by a three-level model, which agrees with the ansatz that a molecule whose nonlinear susceptibility is at the {\em fundamental limit} must be identically a three-level system.  When the hyperpolarizability is below the {\em apparent limit}, usually the oscillator strength is shared between many states, which results in dilution of the hyperpolarizability.  Thus, our calculation agrees with real molecules, which are all measured to be below the {\em apparent limit} and require many states in order to model the nonlinear susceptibility.  Ironically, for parameters that make our simple molecules describable by a two-state system, the value of $\beta$ is found to be well below the {\em fundamental limit}.

To summarize, our simple symmetry-based calculations suggests the ideal geometry and charge distribution of an asymmetric octuple that maximizes the hyperpolarizability, confirms the three-level ansatz that molecules with $\beta$ near the {\em fundamental limit} are dominated by three excited states, and may shed light on the origin of the gap between the best molecules and the {\em fundamental limit}.

\section{Acknowledgements}

MGK thanks The National Science Foundation (ECS-0354736) and Wright-Paterson Air Force Base for generously supporting this work.

\begin{thebibliography}{27}
\expandafter\ifx\csname natexlab\endcsname\relax\def\natexlab#1{#1}\fi
\expandafter\ifx\csname bibnamefont\endcsname\relax
  \def\bibnamefont#1{#1}\fi
\expandafter\ifx\csname bibfnamefont\endcsname\relax
  \def\bibfnamefont#1{#1}\fi
\expandafter\ifx\csname citenamefont\endcsname\relax
  \def\citenamefont#1{#1}\fi
\expandafter\ifx\csname url\endcsname\relax
  \def\url#1{\texttt{#1}}\fi
\expandafter\ifx\csname urlprefix\endcsname\relax\def\urlprefix{URL }\fi
\providecommand{\bibinfo}[2]{#2}
\providecommand{\eprint}[2][]{\url{#2}}

\bibitem[{\citenamefont{Orr and Ward}(1971)}]{orr71.01}
\bibinfo{author}{\bibfnamefont{B.~J.} \bibnamefont{Orr}} \bibnamefont{and}
  \bibinfo{author}{\bibfnamefont{J.~F.} \bibnamefont{Ward}},
  \bibinfo{journal}{Molecular Physics} \textbf{\bibinfo{volume}{20}},
  \bibinfo{pages}{513} (\bibinfo{year}{1971}).

\bibitem[{\citenamefont{Kuzyk}(2000{\natexlab{a}})}]{kuzyk00.01}
\bibinfo{author}{\bibfnamefont{M.~G.} \bibnamefont{Kuzyk}},
  \bibinfo{journal}{Phys. Rev. Lett.} \textbf{\bibinfo{volume}{85}},
  \bibinfo{pages}{1218} (\bibinfo{year}{2000}{\natexlab{a}}).

\bibitem[{\citenamefont{Kuzyk}(2003{\natexlab{a}})}]{kuzyk03.02}
\bibinfo{author}{\bibfnamefont{M.~G.} \bibnamefont{Kuzyk}},
  \bibinfo{journal}{Phys. Rev. Lett.} \textbf{\bibinfo{volume}{90}},
  \bibinfo{pages}{039902} (\bibinfo{year}{2003}{\natexlab{a}}).

\bibitem[{\citenamefont{Kuzyk}(2000{\natexlab{b}})}]{kuzyk00.02}
\bibinfo{author}{\bibfnamefont{M.~G.} \bibnamefont{Kuzyk}},
  \bibinfo{journal}{Opt. Lett.} \textbf{\bibinfo{volume}{25}},
  \bibinfo{pages}{1183} (\bibinfo{year}{2000}{\natexlab{b}}).

\bibitem[{\citenamefont{Kuzyk}(2003{\natexlab{b}})}]{kuzyk03.01}
\bibinfo{author}{\bibfnamefont{M.~G.} \bibnamefont{Kuzyk}},
  \bibinfo{journal}{Opt. Lett.} \textbf{\bibinfo{volume}{28}},
  \bibinfo{pages}{135} (\bibinfo{year}{2003}{\natexlab{b}}).

\bibitem[{\citenamefont{Olbrechts et~al.}(2000)\citenamefont{Olbrechts, Clays,
  Wostyn, and Persoons}}]{olbre20.01}
\bibinfo{author}{\bibfnamefont{G.}~\bibnamefont{Olbrechts}},
  \bibinfo{author}{\bibfnamefont{K.}~\bibnamefont{Clays}},
  \bibinfo{author}{\bibfnamefont{K.}~\bibnamefont{Wostyn}}, \bibnamefont{and}
  \bibinfo{author}{\bibfnamefont{A.}~\bibnamefont{Persoons}},
  \bibinfo{journal}{Synthetic Metals} \textbf{\bibinfo{volume}{115}},
  \bibinfo{pages}{207} (\bibinfo{year}{2000}).

\bibitem[{\citenamefont{Kuzyk}(2001)}]{kuzyk01.01}
\bibinfo{author}{\bibfnamefont{M.~G.} \bibnamefont{Kuzyk}},
  \bibinfo{journal}{IEEE Journal on Selected Topics in Quantum Electronics}
  \textbf{\bibinfo{volume}{7}}, \bibinfo{pages}{774 } (\bibinfo{year}{2001}).

\bibitem[{\citenamefont{Clays}(2001)}]{clays01.01}
\bibinfo{author}{\bibfnamefont{K.}~\bibnamefont{Clays}}, \bibinfo{journal}{Opt.
  Lett.} \textbf{\bibinfo{volume}{26}}, \bibinfo{pages}{1699}
  (\bibinfo{year}{2001}).

\bibitem[{\citenamefont{Kuzyk}(2003{\natexlab{c}})}]{kuzyk03.03}
\bibinfo{author}{\bibfnamefont{M.~G.} \bibnamefont{Kuzyk}},
  \bibinfo{journal}{J. Chem Phys.} \textbf{\bibinfo{volume}{119}}
  (\bibinfo{year}{2003}{\natexlab{c}}).

\bibitem[{\citenamefont{Perez-Moreno and Kuzyk}(2005)}]{perez05.01}
\bibinfo{author}{\bibfnamefont{X.}~\bibnamefont{Perez-Moreno}}
  \bibnamefont{and} \bibinfo{author}{\bibfnamefont{M.~G.} \bibnamefont{Kuzyk}},
  \bibinfo{journal}{J. Chem. Phys.} \textbf{\bibinfo{volume}{123}},
  \bibinfo{pages}{194101} (\bibinfo{year}{2005}).

\bibitem[{\citenamefont{Kuzyk}(2004)}]{kuzyk04.02}
\bibinfo{author}{\bibfnamefont{M.~G.} \bibnamefont{Kuzyk}},
  \bibinfo{journal}{J. Nonl. Opt. Phys. \& Mat.} \textbf{\bibinfo{volume}{13}},
  \bibinfo{pages}{461} (\bibinfo{year}{2004}).

\bibitem[{\citenamefont{Kuzyk}(2003{\natexlab{d}})}]{Kuzyk03.04}
\bibinfo{author}{\bibfnamefont{M.~G.} \bibnamefont{Kuzyk}},
  \bibinfo{journal}{IEEE Circuits and Devices Magazine}
  \textbf{\bibinfo{volume}{19}}, \bibinfo{pages}{8}
  (\bibinfo{year}{2003}{\natexlab{d}}).

\bibitem[{\citenamefont{May et~al.}(2005)\citenamefont{May, Lim, Biaggio,
  Moonen, Michinobu, and Diederich}}]{May05.01}
\bibinfo{author}{\bibfnamefont{J.~C.} \bibnamefont{May}},
  \bibinfo{author}{\bibfnamefont{J.~H.} \bibnamefont{Lim}},
  \bibinfo{author}{\bibfnamefont{I.}~\bibnamefont{Biaggio}},
  \bibinfo{author}{\bibfnamefont{N.~N.~P.} \bibnamefont{Moonen}},
  \bibinfo{author}{\bibfnamefont{T.}~\bibnamefont{Michinobu}},
  \bibnamefont{and}
  \bibinfo{author}{\bibfnamefont{F.}~\bibnamefont{Diederich}},
  \bibinfo{journal}{Opt. Lett.} \textbf{\bibinfo{volume}{30}},
  \bibinfo{pages}{3057} (\bibinfo{year}{2005}).

\bibitem[{\citenamefont{Slepkov et~al.}(2004)\citenamefont{Slepkov, Hegmann,
  Eisler, Elliot, and Tykwinski}}]{slepk04.01}
\bibinfo{author}{\bibfnamefont{A.~D.} \bibnamefont{Slepkov}},
  \bibinfo{author}{\bibfnamefont{F.~A.} \bibnamefont{Hegmann}},
  \bibinfo{author}{\bibfnamefont{S.}~\bibnamefont{Eisler}},
  \bibinfo{author}{\bibfnamefont{E.}~\bibnamefont{Elliot}}, \bibnamefont{and}
  \bibinfo{author}{\bibfnamefont{R.~R.} \bibnamefont{Tykwinski}},
  \bibinfo{journal}{J. Chem. Phys.} \textbf{\bibinfo{volume}{120}},
  \bibinfo{pages}{6807} (\bibinfo{year}{2004}).

\bibitem[{\citenamefont{Kuzyk}(2003{\natexlab{e}})}]{Kuzyk03.05}
\bibinfo{author}{\bibfnamefont{M.~G.} \bibnamefont{Kuzyk}},
  \bibinfo{journal}{Optics \& Photonics News} \textbf{\bibinfo{volume}{14}},
  \bibinfo{pages}{26} (\bibinfo{year}{2003}{\natexlab{e}}).

\bibitem[{\citenamefont{Tripathi et~al.}(2004)\citenamefont{Tripathi, Moreno,
  Kuzyk, Coe, Clays, and Kelley}}]{Tripa04.01}
\bibinfo{author}{\bibfnamefont{K.}~\bibnamefont{Tripathi}},
  \bibinfo{author}{\bibfnamefont{P.}~\bibnamefont{Moreno}},
  \bibinfo{author}{\bibfnamefont{M.~G.} \bibnamefont{Kuzyk}},
  \bibinfo{author}{\bibfnamefont{B.~J.} \bibnamefont{Coe}},
  \bibinfo{author}{\bibfnamefont{K.}~\bibnamefont{Clays}}, \bibnamefont{and}
  \bibinfo{author}{\bibfnamefont{A.~M.} \bibnamefont{Kelley}},
  \bibinfo{journal}{J. Chem. Phys.} \textbf{\bibinfo{volume}{121}},
  \bibinfo{pages}{7932} (\bibinfo{year}{2004}).

\bibitem[{\citenamefont{Chen et~al.}(2004)\citenamefont{Chen, Kuang, Wang, and
  Sargent}}]{wang04.01}
\bibinfo{author}{\bibfnamefont{Q.~Y.} \bibnamefont{Chen}},
  \bibinfo{author}{\bibfnamefont{L.}~\bibnamefont{Kuang}},
  \bibinfo{author}{\bibfnamefont{Z.~Y.} \bibnamefont{Wang}}, \bibnamefont{and}
  \bibinfo{author}{\bibfnamefont{E.~H.} \bibnamefont{Sargent}},
  \bibinfo{journal}{Nano. Lett.} \textbf{\bibinfo{volume}{4}},
  \bibinfo{pages}{1673} (\bibinfo{year}{2004}).

\bibitem[{\citenamefont{Kuzyk}(2005{\natexlab{a}})}]{kuzyk05.01}
\bibinfo{author}{\bibfnamefont{M.~G.} \bibnamefont{Kuzyk}},
  \bibinfo{journal}{Phys. Rev. Lett.} \textbf{\bibinfo{volume}{95}},
  \bibinfo{pages}{109402} (\bibinfo{year}{2005}{\natexlab{a}}).

\bibitem[{\citenamefont{Kuzyk}(2005{\natexlab{b}})}]{kuzyk05.02a}
\bibinfo{author}{\bibfnamefont{M.~G.} \bibnamefont{Kuzyk}},
  \bibinfo{journal}{arXiv:physics/0510002}
  (\bibinfo{year}{2005}{\natexlab{b}}).

\bibitem[{\citenamefont{Champagne and Kirtman}(2005)}]{champ05.01}
\bibinfo{author}{\bibfnamefont{B.}~\bibnamefont{Champagne}} \bibnamefont{and}
  \bibinfo{author}{\bibfnamefont{B.}~\bibnamefont{Kirtman}},
  \bibinfo{journal}{Phys. Rev. Lett.} \textbf{\bibinfo{volume}{95}},
  \bibinfo{pages}{109401} (\bibinfo{year}{2005}).

\bibitem[{\citenamefont{Zienkiewicz et~al.}(2005)\citenamefont{Zienkiewicz,
  Taylor, and Zhu}}]{zienk05.01}
\bibinfo{author}{\bibfnamefont{O.~C.} \bibnamefont{Zienkiewicz}},
  \bibinfo{author}{\bibfnamefont{R.~L.} \bibnamefont{Taylor}},
  \bibnamefont{and} \bibinfo{author}{\bibfnamefont{J.~Z.} \bibnamefont{Zhu}},
  \emph{\bibinfo{title}{The Finite Element Method: Its Basis and Fundamentals}}
  (\bibinfo{publisher}{Butterworth-Heinemanm}, \bibinfo{year}{2005}),
  \bibinfo{edition}{6th} ed.

\bibitem[{\citenamefont{Atkinson and Han}(2001)}]{atkin01.01}
\bibinfo{author}{\bibfnamefont{K.}~\bibnamefont{Atkinson}} \bibnamefont{and}
  \bibinfo{author}{\bibfnamefont{W.}~\bibnamefont{Han}},
  \emph{\bibinfo{title}{Theoretical Numerical Analysis, a Functional Analysis
  Framework}} (\bibinfo{publisher}{Springer}, \bibinfo{address}{New York},
  \bibinfo{year}{2001}).

\bibitem[{\citenamefont{Sorensen}(1992)}]{soren92.01}
\bibinfo{author}{\bibfnamefont{D.~C.} \bibnamefont{Sorensen}},
  \bibinfo{journal}{{SIAM} J. Matrix Anal. Appl.}
  \textbf{\bibinfo{volume}{13}}, \bibinfo{pages}{357} (\bibinfo{year}{1992}).

\bibitem[{\citenamefont{Lehoucq et~al.}(1998)\citenamefont{Lehoucq, Sorensen,
  and Yang}}]{lehou98.01}
\bibinfo{author}{\bibfnamefont{R.~B.} \bibnamefont{Lehoucq}},
  \bibinfo{author}{\bibfnamefont{D.~C.} \bibnamefont{Sorensen}},
  \bibnamefont{and} \bibinfo{author}{\bibfnamefont{C.}~\bibnamefont{Yang}},
  \emph{\bibinfo{title}{ARPACK Users' Guide: Solution of Large-Scale Eigenvalue
  Problems with Implicitly Restarted Arnoldi Methods}}
  (\bibinfo{publisher}{SIAM}, \bibinfo{address}{Philadelphia},
  \bibinfo{year}{1998}).

\bibitem[{\citenamefont{Kuzyk}(2005{\natexlab{c}})}]{kuzyk05.01a}
\bibinfo{author}{\bibfnamefont{M.~G.} \bibnamefont{Kuzyk}},
  \bibinfo{journal}{arXiv:physics/0505006}
  (\bibinfo{year}{2005}{\natexlab{c}}).

\bibitem[{\citenamefont{Kuzyk}(2005{\natexlab{d}})}]{kuzyk05.02}
\bibinfo{author}{\bibfnamefont{M.~G.} \bibnamefont{Kuzyk}},
  \bibinfo{journal}{Phys. Rev. A} \textbf{\bibinfo{volume}{72}},
  \bibinfo{pages}{053819} (\bibinfo{year}{2005}{\natexlab{d}}).

\bibitem[{\citenamefont{Joffre et~al.}(1992)\citenamefont{Joffre, Yaron,
  Silbey, and Zyss}}]{Joffr92.01}
\bibinfo{author}{\bibfnamefont{M.}~\bibnamefont{Joffre}},
  \bibinfo{author}{\bibfnamefont{D.}~\bibnamefont{Yaron}},
  \bibinfo{author}{\bibfnamefont{J.}~\bibnamefont{Silbey}}, \bibnamefont{and}
  \bibinfo{author}{\bibfnamefont{J.}~\bibnamefont{Zyss}}, \bibinfo{journal}{J .
  Chem. Phys.} \textbf{\bibinfo{volume}{97}}, \bibinfo{pages}{5607}
  (\bibinfo{year}{1992}).

\end{thebibliography}

\end{document}